\documentclass[prl,twocolumn,showpacs,preprintnumbers,amsmath,amssymb,noshowpacs,superscriptaddress]{revtex4}
\usepackage{graphicx}
\usepackage{dcolumn}
\usepackage{bm}

\begin{document}

\title{Approaching the Dirac point in high mobility multilayer epitaxial graphene}
\author{M. Orlita}
\email{orlita@karlov.mff.cuni.cz} \affiliation{Grenoble High Magnetic Field
Laboratory, CNRS, BP 166, F-38042 Grenoble Cedex 09, France}
\affiliation{Institute of Physics, Charles University, Ke Karlovu 5, CZ-121~16
Praha 2, Czech Republic} \affiliation{Institute of Physics, v.v.i., ASCR,
Cukrovarnick\'{a} 10, CZ-162 53 Praha 6, Czech Republic}
\author{C. Faugeras}
\affiliation{Grenoble High Magnetic Field Laboratory, CNRS, BP 166, F-38042
Grenoble Cedex 09, France}
\author{P. Plochocka}
\affiliation{Grenoble High Magnetic Field Laboratory, CNRS, BP 166, F-38042
Grenoble Cedex 09, France}
\author{P. Neugebauer}
\affiliation{Grenoble High Magnetic Field Laboratory, CNRS, BP 166, F-38042
Grenoble Cedex 09, France}
\author{G. Martinez}
\affiliation{Grenoble High Magnetic Field Laboratory, CNRS, BP 166, F-38042
Grenoble Cedex 09, France}
\author{D. K. Maude}
\affiliation{Grenoble High Magnetic Field Laboratory, CNRS, BP 166, F-38042
Grenoble Cedex 09, France}
\author{A.-L. Barra}
\affiliation{Grenoble High Magnetic Field Laboratory, CNRS, BP 166, F-38042
Grenoble Cedex 09, France}
\author{M. Sprinkle}
\affiliation{School of Physics, Georgia Institute of Technology, Atlanta,
Georgia 30332, USA}
\author{C. Berger}
\affiliation{School of Physics, Georgia Institute of Technology, Atlanta,
Georgia 30332, USA} \affiliation{Institut N\'{e}el/CNRS-UJF BP 166,
F-38042 Grenoble Cedex 9, France}
\author{W. A. de Heer}
\affiliation{School of Physics, Georgia Institute of Technology, Atlanta,
Georgia 30332, USA}
\author{M. Potemski}
\affiliation{Grenoble High Magnetic Field Laboratory, CNRS, BP 166, F-38042
Grenoble Cedex 09, France}

\date{\today}

\begin{abstract}
Multi-layer epitaxial graphene (MEG) is investigated using far
infrared (FIR) transmission experiments in the different limits of low
magnetic fields and high temperatures. The cyclotron-resonance like
absorption is observed at low temperature in magnetic fields below
50~mT, allowing thus to probe the nearest vicinity of the Dirac
point and to estimate the conductivity in nearly~undoped graphene.
The carrier mobility is found to exceed 250,000 cm$^2$/(V.s). In the
limit of high temperatures, the well-defined Landau level (LL)
quantization is observed up to room temperature at magnetic fields
below 1 T, a phenomenon unique in solid state systems. A negligible
increase in the width of the cyclotron resonance lines with
increasing temperature indicates that no important scattering
mechanism is thermally activated, supporting recent expectations of
high room-temperature mobilities in graphene.
\end{abstract}

\pacs{71.70.Di, 76.40.+b, 78.30.-j, 81.05.Uw}

\maketitle

The quality of electronic systems, measured in terms of the carrier
mobility, and determined by carrier scattering processes, is a major
issue in material science both from a fundamental physics and an
applications point of view. This issue is now critical for further
progress in the research on 2D allotropes of
carbon~\cite{BergerJPCB04,NovoselovNature05,ZhangNature05,BergerScience06,GeimNatureMaterial07,CastroNetoRMP08}.
In exfoliated graphene, charged impurities are recognized as the
dominant source of scattering in transport
experiments~\cite{AndoJPSJ06,NomuraPRL07,HwangPRL07}. The presence
of charged impurities even prevents experiments in the vicinity of
the Dirac point, due to the electron and hole puddles which
inevitably emerge in these samples when lowering
density~\cite{MartinNaturePhysics07}. Recently, the first
experiments on suspended graphene~\cite{BolotinSSC08,DuNN08}, showed
a promising way to come closer to Dirac point, but nevertheless, the
exact role of individual scattering mechanisms, especially close to
Dirac point, together with the nature of the minimum conductivity
remains
unclear~\cite{ShonJPSJ98,TanPRL07,AdamPNAS07,CheianovPRL07,ChenNaturePhys08}.

\begin{figure}
\scalebox{0.47}{\includegraphics*[angle=270,bb= 123pt 61pt 453pt
475pt]{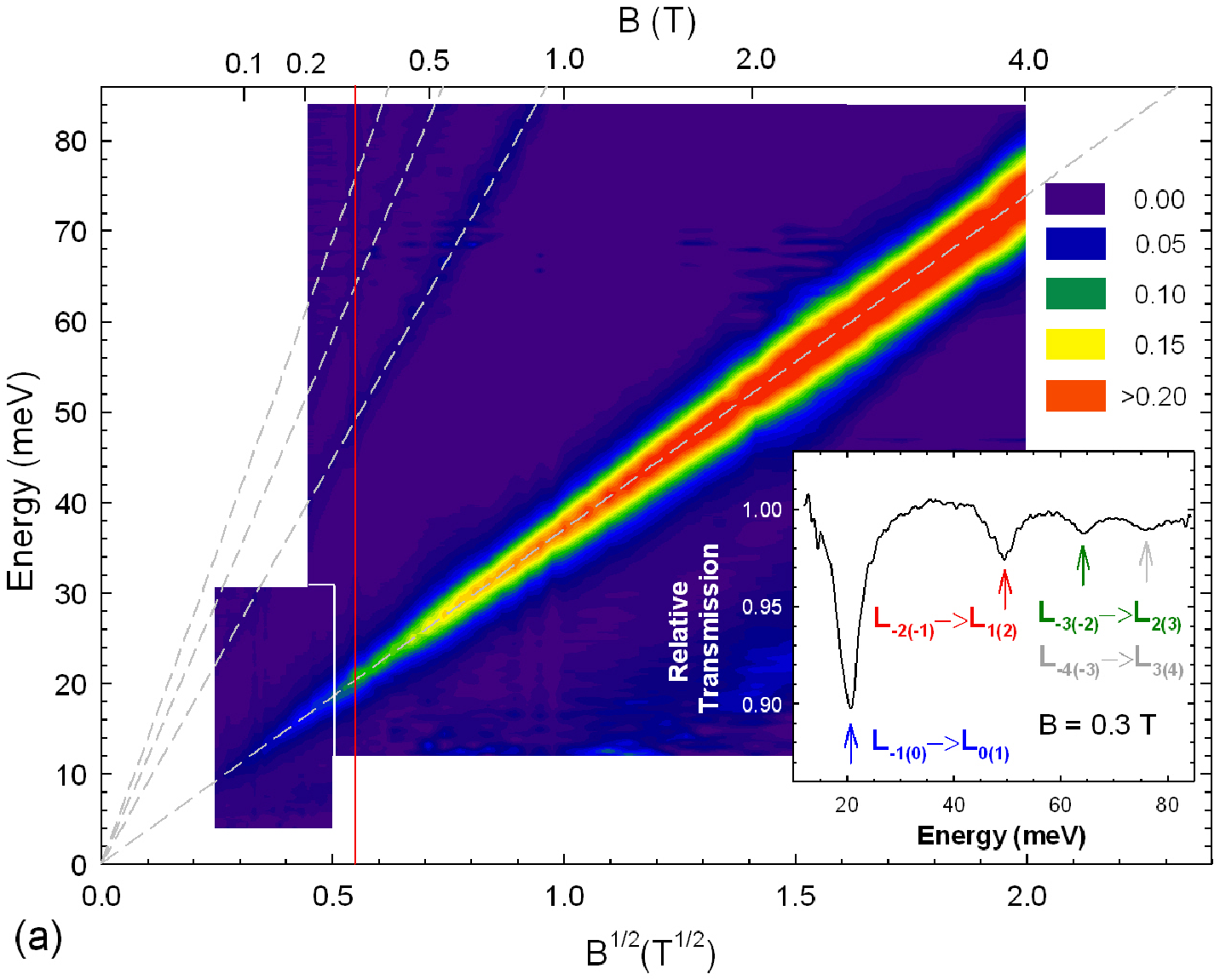}}
\scalebox{0.95}{\includegraphics*[19pt,17pt][241pt,158pt]{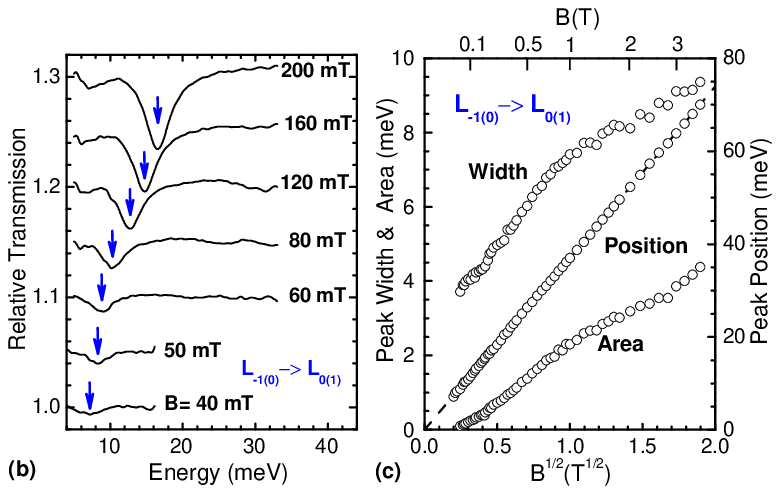}}
\caption{\label{SPKTLowTemp} (color online) Part (a): The FIR transmission
$\mathcal{T}$ plotted as $-\ln{\mathcal{T}}$  as a function of the magnetic
field at $T=2.0$~K. The dashed lines denote the expected transitions for
$\tilde{c}=1.02\times10^{6}$~m.s$^{-1}$. The inset shows the transmission
spectrum at $B=0.3$~T. Part (b): FIR transmission  taken at $T=2$~K in low
magnetic fields. For clarity, successive spectra are shifted vertically by
0.05. The part (c) shows the peak position, width and area for the
L$_{-1(0)}\rightarrow$L$_{0(1)}$ transition. The dashed line in part (c) is a
least squares fit to the peak positions.}
\end{figure}

In a parallel track to the exfoliated-graphene research, the
equivalent problems arise in the study of epitaxial graphene.
Multi-layer epitaxial graphene (MEG) can be relatively easily
thermally decomposed from SiC on the C-terminated surface and
consist of electrically decoupled graphene
layers~\cite{HassPRB07,HassPRL08}. It is highly conducting in the
very close vicinity of the interface with SiC, as a result of charge
transfer from SiC~\cite{BergerScience06,VarchonPRL08}. The
subsequent layers are practically neutral~\cite{SadowskiPRL06}.
Electrical measurements on epitaxial graphene samples probe mostly
the low resistance interface layers. However, the quasi-neutral part
of the MEG can be investigated by FIR optical absorption technique
in an applied magnetic field~\cite{SadowskiPRL06}.

Both epitaxial and exfoliated graphene have been extensively
investigated at elevated temperatures. By analogy with room
temperature ballistic transport in carbon
nanotubes~\cite{FrankScience98}, very high mobilities are expected
for graphene at room temperature. This is indeed suggested by the
investigation of the temperature dependence of the extrinsic and
intrinsic scattering mechanisms in exfoliated
graphene~\cite{MorozovPRL08} and in the weak temperature dependence
of the resistance in epitaxial graphene~\cite{BergerScience06}.

In this Letter, we focus on carrier scattering processes in the
majority graphene layers with a low carrier concentration and show
their extraordinary quality, expressed by the temperature
independent carrier mobility in excess of 250,000~cm$^2$/(V.s). This
information is extracted from the analysis of inter-LL transitions
which are investigated in the relevant limits of high temperatures
and low magnetic fields.

The investigated graphene sample was prepared by thermal
decomposition from a 4H-SiC substrate~\cite{BergerJPCB04} and
contains around $\sim$100 graphene layers. The sample was
characterized using micro-Raman experiment, which revealed similarly
to Ref.~\cite{FaugerasAPL08} the presence of decoupled graphene
layers as well as of additional graphite residuals. The Raman
experiment with the laser spot of diameter $\sim$3~$\mu$m shows an
extraordinarily narrow 2D band ($\sim$20~cm$^{-1}$) compared to
values reported on exfoliated graphene~\cite{FerrariPRL06}. To
measure the FIR transmittance of the sample, the macroscopic area of
the sample ($\sim$4~mm$^2$) was exposed to the radiation of a globar
or mercury lamp, which was analyzed by a Fourier transform
spectrometer and delivered to the sample via light-pipe optics. In
experiments performed at $T=2.0$~K, the light was detected by a
composite bolometer placed directly below the sample, while at
higher temperatures an external bolometer was used. All presented
transmission spectra are normalized by the sample transmission at
$B=0$.

The low temperature FIR transmission spectra of the investigated
sample, shown in Fig.~\ref{SPKTLowTemp}, are typical of the optical
response of MEG~\cite{SadowskiPRL06}. Considering the LL spectrum in
graphene:
$E_n=\mathrm{sign}(n)\tilde{c}\sqrt{2e\hbar B|n|}$, four absorption
lines, denoted by arrows in the inset of Fig.~\ref{SPKTLowTemp}(a)
and showing a $\sqrt{B}$-scaled blueshift, can be identified as
inter-LL transitions L$_{-m}\rightarrow$L$_{m+1}$ and
L$_{-(m+1)}\rightarrow$L$_m$ with $m=0,1,2$ and 3. The corresponding
Fermi velocity was evaluated as $\tilde{c}=(1.02\pm 0.01)\times
10^{6}$~m.s$^{-1}$. No deviations from the single particle model due
to many-body effects are found in MEG in contrast to recent
experiments performed on exfoliated
samples~\cite{JiangPRL07,LiNaturePhys08}. Notably, the spectra of
this $\sim$100 layer sample present no transitions symmetric around
Dirac point (L$_{-m}\rightarrow$L$_m$) which are characteristic of
bulk graphite~\cite{OrlitaPRL08}.

Henceforward, we focus on the main line in the spectra,
L$_{0(-1)}\rightarrow$L$_{1(0)}$, which necessarily corresponds to
transitions from or to the vicinity of the Fermi level. Following
this transition with the magnetic field, we find no deviation from
the $\sqrt{B}$-scaling and the line is still visible at $B\sim
40$~mT when it is centered at an energy of $\approx$7~meV. The FIR
experiment, thus allows to probe the very close vicinity of Dirac
point, hardly accessible in the current transport
experiments~\cite{BergerScience06,BolotinSSC08,DuNN08}, and also
shows that the linearity of the density of states is preserved down
to the distances of few meV from Dirac point. The disappearance of
the line below $B\sim 40$~mT gives an estimation of the LL filling
factor $\nu \approx 6$ at this field with the corresponding carrier
density of $n_0\approx 5\times 10^{9}$~cm$^{-2}$. The possible
remanent field of the solenoid of $\leq5$~mT limits the accuracy of
this estimation to 10\%. Note that this density is about three
orders of magnitude smaller than the carrier density measured in
transport for equivalent
samples~\cite{BergerJPCB04,BergerScience06,WuPRL07}, but electrical
conductance is governed by highly doped graphene layer(s) close to
the SiC substrate. While the FIR experiment probes all layers in the
sample, the strongly doped layer(s) give no contribution to the FIR
spectra in the presented region of energies and magnetic fields.

The shape of the main line can be well reproduced by a simple
Lorentzian curve. The results of the fitting procedure are shown in
Fig.~\ref{SPKTLowTemp}(c), where the peak position, area and width
are plotted as a function of $\sqrt{B}$. Both the peak position and
the area show a linear increase with $\sqrt{B}$, in agreement with
expectations for a single graphene layer~\cite{SadowskiPRL06}. The
latter also suggests a relatively good homogeneity of the carrier
density, as the significant presence of more doped regions would
result in a superlinear rise. An interesting evolution with magnetic
field is seen for the linewidth $\delta E$. Starting from $\delta E
\sim 4$~meV at the lowest magnetic field $B=60$~mT, where the
line-shape analysis is possible, the width increases nearly linearly
with $\sqrt{B}$ up to almost $8$~meV at $1$~T. At higher magnetic
fields, the increase continues but as a sublinear function of
$\sqrt{B}$. The observed broadening cannot result from the
electron-hole asymmetry, where a broadening which varies linearly
with $B$ is expected~\cite{PlochockaPRL08}. The $\sqrt{B}$-dependent
broadening of LLs was suggested by Shon and Ando~\cite{ShonJPSJ98}
for the case of both short- and long-range scatterers, whose
strength is independent of carrier density. Whereas the short-range
scatterers should induce an identical broadening of all LLs, the
$n=0$ LL should be broadened by a factor of $\sqrt{2}$ more than
other LLs for long-range scatterers. As the width of the main line
is not enhanced compared to the other transitions, we can conclude
that short-range scattering is probably dominant here.

\begin{figure}
\scalebox{1.2}{\includegraphics*[18pt,17pt][170pt,199pt]{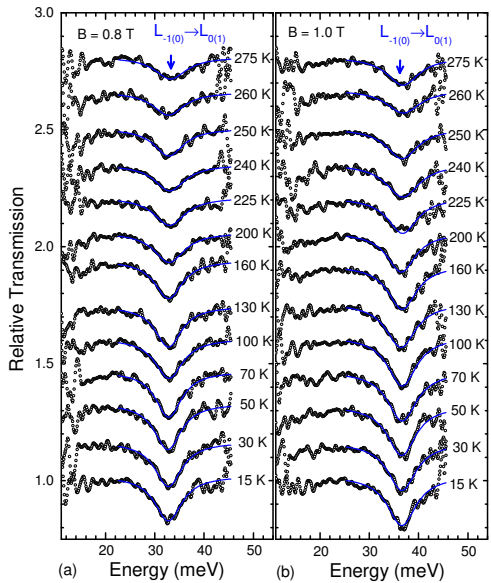}}
\caption{\label{SPKT}  (color online) Temperature dependence of the
transmission spectra taken at magnetic fields $B=0.8$ and 1.0~T in parts (a)
and (b), respectively, showing the L$_{-1(0)}\rightarrow$L$_{0(1)}$ transition
up to the room temperature. The blue curves represent the Lorentzian fits whose
parameters are plotted in Fig.~\ref{Analysis}. For clarity, successive spectra
in both parts are shifted vertically by 0.15.}
\end{figure}

To compare our results with the recent transport experiments, a
simple estimation of the scattering time, mobility and conductivity
are presented. From the width of the main transition $\delta E$, the
carrier scattering time can be estimated using $\tau=2\hbar/\delta
E$, also used to compare FIR transmission and transport experiments
performed on exfoliated graphene~\cite{JiangPRL07,HenriksenPRL08}.
The scattering time, obtained for our macroscopic sample, is
$\tau\sim 300$~fs at low magnetic fields ($\delta E \approx 4$~meV)
and decreases with increasing $B$ to $\tau\sim 150$~fs at $B=1$~T
($\delta E \approx 8$~meV). Equivalent scattering times
$\tau\sim200$~fs are nowadays reported for single-flake graphene on
a Si/SiO$_{2}$ substrate at densities around $10^{12}$~cm$^{-2}$ but
this time decreases rapidly with decreasing carrier
density~\cite{TanPRL07}. The scattering time of $\tau\sim260$~fs was
also reported in charged graphene layer ($4\times10^{12}$~cm$^{-2}$)
at SiC/graphene interface~\cite{WuPRL07}. Recently, scattering times
$\tau\sim100$~fs were achieved at density down to
$\sim10^{10}$~cm$^{-2}$ in suspended
graphene~\cite{BolotinSSC08,DuNN08}.

The lowest field of $B=40$~mT for which the well-defined absorption
line is observed at $\hbar\omega_{c}\approx7$~meV allows an
independent estimation of the lower bound for $\tau$, and the
mobility $\mu$. Using the semi-classical condition $\omega_c\tau>1$,
allowing the carriers to complete one cyclotron orbit without
scattering, gives a minimum relaxation time
$\tau>\hbar/$7~meV$\simeq100$~fs independent of but in a good
agreement with the previous estimation based on the linewidth
$\delta E$. The condition $\omega_c\tau>1$ can be rewritten as $\mu
B>1$, which leads to the mobility $\mu > 0.25\times
10^6$~cm$^{2}$/(V.s). Note that this value represents only the
lowest bound, as the main absorption line disappears from the
spectrum due to the complete filling (depopulation) of $n=1$
($n=-1$) LL and not due to the LL broadening. It remains an open
question how this mobility would evolve with increasing carrier
density, since the scattering mechanisms are different from those in
exfoliated graphene, where ionized impurities dominate and the
mobility remains nearly
constant~\cite{TanPRL07,BolotinSSC08,ChenNaturePhys08,DuNN08}.

To estimate the conductivity, we can use the standard relation
$\sigma(\varepsilon)=(e^2/2)D(\varepsilon)\tilde{c}^2\tau$, derived
in the Boltzmann transport theory. This formula is applicable for
graphene systems with a homogeneous density of
carriers~\cite{HwangPRL07,NomuraPRL07}, fulfilled in our case, but
distinctive deviations from a presumably more precise
self-consistent Born approximation are predicted in the immediate
vicinity of Dirac point~\cite{ShonJPSJ98}. To calculate the density
of states $D(\varepsilon)$, we recall that no deviation from the
$\sqrt{B}$ scaling of the position of the main line is observed, and
thus the expression $D(\varepsilon)=g_sg_v
|\varepsilon|/(2\pi\tilde{c}^2\hbar^2)$ remains valid down to a few
meV from Dirac point, where $g_s=g_v=2$ denote the spin and valley
degeneracy, respectively. Finally, we obtain the conductivity in the
form $\sigma(\varepsilon)=(2e^2/h)|\varepsilon|(\tau/\hbar)$. In the
limit of a very low magnetic field, when the LL quantization energy
is comparable to the LL broadening, we find $\delta
E=2\hbar/\tau\rightarrow \approx 3$~meV and the Fermi level
$|\varepsilon|\rightarrow\tilde{c}\hbar\sqrt{\pi n_0}\approx 8$~meV
which implies a zero-field conductivity
$\sigma=(4e^2/h)(|\varepsilon|/\delta E) \approx 10e^2/h$, slightly
larger than the minimum conductivity reported in
Refs.~\cite{NovoselovNature05,ZhangNature05,TanPRL07}, but
corresponding to conclusions for relatively clean
samples~\cite{AdamPNAS07,ChenNaturePhys08}.

\begin{figure}
\scalebox{1.8}{\includegraphics*[12pt,10pt][85pt,131pt]{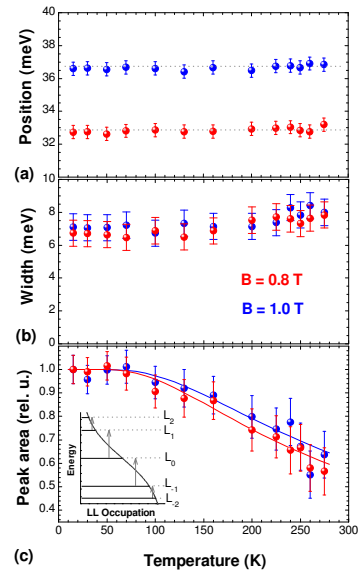}}
\caption{\label{Analysis}  (color online) Parameters of the
Lorentzian fits to the transmission spectra in Fig.~\ref{SPKT}. Red
and blue color represent data for $B=0.8$ and 1.0~T, respectively.
Whereas position and width of absorption lines in parts (a) and (b)
show none or a very a weak temperature dependence, the peak area in
the part (c) significantly decreases with T, nearly following the
theoretical expectation $1-2f_1$ depicted by solid lines. The inset
schematically shows the LL occupation and possible dipole-allowed
absorption processes between LLs at finite temperature in undoped
graphene ($E_F=0$).}
\end{figure}

Having discussed the spectra taken at $T=2.0$~K, we turn our
attention to measurements at elevated temperatures in
Fig.~\ref{SPKT}. In these data, we can follow the evolution of the
L$_{-1(0)}\rightarrow$L$_{0(1)}$ transitions at $B=0.8$ and 1.0~T in
the interval of temperatures $T=15-275$~K. Apart from an apparent
decrease in the absorption line intensity, no other effects are
noticeable. Hence, the well-defined LLs are uniquely observed in
graphene at room temperature and for magnetic fields below 1 T, much
lower than reported up to now~\cite{NovoselovScience07}. The blue
curves in Fig.~\ref{SPKT} are Lorentzian fits to the data, and the
obtained parameters, the peak position, width and area are plotted
in Fig.~\ref{Analysis}. Based on these results, we can draw the
following conclusions. The constant position of the absorption line
shows that the Fermi velocity is independent of temperature. The
extremely weak increase of the linewidth shows that no important
scattering mechanism is thermally-activated, indicating that the
mobility $\mu>0.25\times 10^6 $~cm$^2$/(V.s) remains practically
unchanged up to room temperature. For comparison, the mobility in
standard 2D GaAs systems at room temperature does not exceed
10$^4$~cm$^2$/(V.s), being intrinsically limited by electron
scattering on optical phonons~\cite{WalukiewiczPRB84}. Scattering on
neutral centers, independent of carrier momentum and in consequence
raising the temperature independent mobility, is likely the
predominant scattering mechanisms in the investigated graphene
layers which are well separated from the substrate and screened by
highly conducting interface graphene sheets. Such mechanism is
consistent with the above discussed observation of the nearly
$\sqrt{B}$-dependence of the linewidth. For comparison, the
scattering on ionized impurities determines carrier mobility in
exfoliated graphene, which in contrast may lead to a temperature
dependent mobility at high temperature, in case of a non degenerate
gas.

The decrease in the strength of the transition can be understood by considering
the influence of the thermal population and depopulation of LL's on the
probability of the absorption process. The probabilities for the transitions
L$_{0}\rightarrow$L$_{1}$ and L$_{-1}\rightarrow$L$_{0}$ are expressed as
$P_{0\rightarrow1}\propto (f_0-f_1)$ and $P_{-1\rightarrow 0}\propto
(f_{-1}-f_{0})$, respectively, yielding the total probability $P \propto
(f_{-1}-f_1$), where $f_n$ is the occupation of the $n$-th LL. Assuming undoped
graphene ($E_F=0$), the relative decrease in the intensity of the main
absorption line can be written as $I(T)=1-2f_{1}$ which reproduces the data in
Fig.~\ref{Analysis}(c) extremely well. To calculate $f_1$, the optically
determined Fermi velocity $\tilde{c}=1.02\times10^{6}$~m.s$^{-1}$ was used.
Hence, the only pronounced effect observed in our FIR spectra with increasing
temperature is a population effect due to the thermal excitation of carriers,
as schematically shown in the inset of Fig.~\ref{Analysis}(c).

In summary, MEG has been investigated using LL spectroscopy in low
magnetic fields and at temperatures ranging from 2 to 275~K.
Well-defined inter-LL absorption are observed at low temperatures
down to magnetic fields of 40~mT. The LL spectroscopy thus probes
the electronic states in the immediate vicinity of Dirac point,
which are hardly accessible in experiments on exfoliated
graphene~\cite{TanPRL07,ChenNaturePhys08,BolotinSSC08,DuNN08} due to
the inevitable presence of electron-hole puddles. The conductivity
of nearly neutral MEG deduced from our optical data is found to be
consistent with recent theory~\cite{AdamPNAS07,HwangPRL07}. The
temperature independent width of the spectral lines, which arise
from the temperature independent LL broadening, indicate very weak
thermally activated scattering processes. Therefore, the carrier
mobility in excess of 250,000 cm$^2$/(V.s), evaluated from low
temperature data, remains practically constant with increasing
temperature, confirming the recent expectations of record
room-temperature mobilities in graphene~\cite{MorozovPRL08}. MEG is
an interesting material to probe the properties of Dirac particles
in the immediate vicinity of Dirac point and has a potential as a
platform for high performance graphene-based electronics. However,
realization of any electric transport experiments on nearly neutral
MEG will require passivation of the highly conducting graphene
interface layers. Encouraging research along these lines is already
underway~\cite{WuPRL08}.

\begin{acknowledgments}
The present work was supported by Contract No. ANR- 06-NANO-019, Projects No.
MSM0021620834 and No. KAN400100652, and by the European Commission through
Grant No. RITA-CT-2003-505474. Support is acknowledged from NSF-NIRT 4106A68
and NSF-MRI 4106A95 grants and from  the W. M. Keck foundation. A USA-France
travel grant from CNRS is acknowledged.
\end{acknowledgments}


\end{document}